%% file: Trnsf_based_EmoRec.tex
\documentclass{article}

\usepackage{PRIMEarxiv}

\usepackage[utf8]{inputenc} %
\usepackage[T1]{fontenc}    %
\usepackage{hyperref}       %
\usepackage{url}            %
\usepackage{booktabs}       %
\usepackage{amsfonts}       %
\usepackage{nicefrac}       %
\usepackage{microtype}      %
\usepackage{lipsum}
\usepackage{fancyhdr}       %
\usepackage{graphicx}       %

\usepackage{booktabs}
\usepackage{array}
\usepackage{multirow}
\usepackage[T1]{fontenc}
\usepackage[inline]{enumitem}
\newcolumntype{M}[1]{>{\centering\arraybackslash}m{#1}}

\usepackage{tikz}
\usetikzlibrary{arrows.meta}
\usetikzlibrary{decorations.pathreplacing}
\usetikzlibrary{decorations}

\usepackage{pgfplots}
\pgfplotsset{compat=1.9}

\usepackage{bm}

\usepackage{amsmath}
\usepackage{amsfonts}
\usepackage{hyperref}
\hypersetup{
pdftitle={Transformer-Based Self-Supervised Learning for Emotion Recognition},
pdfsubject={Author Version},
pdfauthor={Juan Vazquez-Rodriguez, Grégoire Lefebvre, Julien Cumin and James L. Crowley},
pdfkeywords={Emotion recognition, machine learning, unsupervised learning, neural networks}
}

\title{Transformer-Based Self-Supervised Learning for Emotion Recognition }

\author{Juan Vazquez-Rodriguez\textsuperscript{1,2},
Grégoire Lefebvre\textsuperscript{1},
Julien Cumin\textsuperscript{1} and
James L. Crowley\textsuperscript{2} \\
\textsuperscript{1}Orange Labs, Grenoble, France\\
Email: \{juan.vazquezrodriguez, gregoire.lefebvre, julien1.cumin\}@orange.com \\
\textsuperscript{2}Univ. Grenoble Alpes, CNRS, Inria, Grenoble INP, LIG, Grenoble, France\\
Email: james.crowley@inria.fr
}

\begin{document}

\maketitle

\begin{abstract}
In order to exploit representations of time-series signals, such as physiological signals, it is essential that these representations capture relevant information from the whole signal. In this work, we propose to use a Transformer-based model to process electrocardiograms (ECG) for emotion recognition. Attention mechanisms of the Transformer can be used to build contextualized representations for a signal, giving more importance to relevant parts. These representations may then be processed with a fully-connected network to predict emotions.

To overcome the relatively small size of datasets with emotional labels, we employ self-supervised learning. We gathered several ECG datasets with no labels of emotion to pre-train our model, which we then fine-tuned for emotion recognition on the AMIGOS dataset. We show that our approach reaches state-of-the-art performances for emotion recognition using ECG signals on AMIGOS. More generally, our experiments show that transformers and pre-training are promising strategies for emotion recognition with physiological signals.

\end{abstract}

\section{Introduction}

When processing time-series signals with deep learning approaches, it is useful to be able to aggregate information from the whole signal, including long-range information, in a way that the most relevant parts are given more importance. One way of doing this is by employing an attention mechanism \cite{attentionPaper} that uses attention weights to limit processing to relevant contextual information, independent of distance. 

Arguably, the Transformer \cite{vaswaniAttentionAllYou2017} is one of the most successful attention-based approaches. Developed for Natural Language Processing (NLP), the Transformer uses attention mechanisms to interpret sequences of words, and is suitable for use in other tasks requiring interpretation of sequences, such as time series forecasting, \cite{liEnhancingLocalityBreaking2019}, analysis of medical physiological signals \cite{yanFusingTransformerModel2019, ahmedt-aristizabalAttentionNetworksMultiTask2020}, and recognition of human activity from motion \cite{haresamudramMaskedReconstructionBased2020}.   

Physiological signal analysis can be seen as a form of time-series analysis and are thus amenable to processing with Transformers. Moreover, these signals can be used to predict emotions \cite{shuReviewEmotionRecognition2018}, and
sensors for these types of signals can be incorporated into wearable devices, as a non-invasive means for monitoring the emotional reaction of users. 
Several works in this direction have emerged using signals like electrocardiograms (ECG) \cite{santamaria-granadosUsingDeepConvolutional2019, sarkarSelfSupervisedLearningECGBased2020}, electroencephalograms (EEG) \cite{songEEGEmotionRecognition2020, beckerEmotionRecognitionBased2020}, electrodermal activity (EDA) \cite{shuklaFeatureExtractionSelection2019}, and other types of physiological signals \cite{chenEmotionRecognitionBased2021, koseNewApproachEmotions2021}.

Established approaches for deep learning with Convolutions and Recurrent networks require large datasets of labeled training data. However, providing ground truth emotion labels for physiological data is a difficult and expensive process, limiting the availability of data for training \cite{koelstraDEAPDatabaseEmotion2012, subramanianASCERTAINEmotionPersonality2018, correaAMIGOSDatasetAffect2018}. 
Pre-training  models with self-supervised learning can help to overcome this lack of labeled training data. With such an approach, during pre-training the model learns general data representations using large volumes of unlabeled data. The model is then fine tuned for a specific task using labeled data. This approach has been successfully used in other domains including NLP \cite{petersDeepContextualizedWord2018, devlinBERTPretrainingDeep2019} and Computer Vision \cite{chenSimpleFrameworkContrastive2020, wuVisualTransformersTokenbased2020}. It has also been successfully used in affective computing, in tasks like emotion recognition from physiological signals \cite{sarkarSelfSupervisedLearningECGBased2020, rossUnsupervisedMultimodalRepresentation2021} and from speech \cite{macaryUseSelfSupervisedPreTrained2021}, personality recognition \cite{songSelfsupervisedLearningPersonspecific2021}, and facial expression recognition \cite{ wilesSelfsupervisedLearningFacial2018, liSelfSupervisedRepresentationLearning2019, roySelfsupervisedContrastiveLearning2021}.

In this paper, we address the problem of predicting emotions from ECG signals. We are interested in obtaining contextualized representations from these signals using a Transformer-based architecture, and then using these representations to predict low/high levels of arousal and valence. We believe that the contextualized representations obtained with the Transformer should capture relevant information from the whole signal, which the performance of the downstream task of emotion recognition should benefit from. 
Our main contributions are: 
\begin{enumerate*}
    \item We show that it is feasible to use a Transformer-based architecture for emotion prediction from ECG signals.
    \item We show that using a self-supervised technique to pre-train the model is useful for ECG signals, achieving superior performance in emotion recognition than a fully-supervised approach. 
    \item We show that our pre-trained Transformer-based model reaches state-of-the-art performances on a dataset of the literature.
\end{enumerate*}

\section{Related Work}

Traditional techniques for emotion recognition from physiological signals include Gaussian naive Bayes, Support Vector Machines, k-Nearest Neighbours, and Random Forests. \cite{subramanianASCERTAINEmotionPersonality2018, correaAMIGOSDatasetAffect2018, gjoreskiInterdomainStudyArousal2018, ayataEmotionRecognitionMultimodal2020, hsuAutomaticECGBasedEmotion2020, shuWearableEmotionRecognition2020}. These approaches typically use manually-selected time and frequency features derived from intuition and domain knowledge. Shukla et al. \cite{shuklaFeatureExtractionSelection2019} show that commonly used features for arousal and valence prediction are not necessarily the most discriminant. This illustrates the difficulty of selecting good hand-crafted features.

To overcome this, researchers have increasingly used deep learning techniques to extract features from physiological signals for emotion recognition. A common approach, described by Santamaria et al. \cite{santamaria-granadosUsingDeepConvolutional2019}, is to use a 1D Convolutional Neural Network (CNN) to extract the features (also called representations), followed by a fully-connected network (FCN) used as classifier to predict emotions. As an alternative, Harper and Southern \cite{harperBayesianDeepLearning2020} use a Long Short-Term Memory (LSTM) network concurrently with a 1D-CNN. Siddharth et al. \cite{siddharthUtilizingDeepLearning2019}, first convert signals into an image using spectrograms \cite{fulopAlgorithmsComputingTimecorrected2006}, and then use a 2D-CNN for feature extraction, followed by an extreme learning machine \cite{huangExtremeLearningMachine2006a} for classification.

One drawback of these CNN-based approaches is that they do not take context into account: after training, kernel weights of the CNN are static, no matter the input. For this reason, attention-based architectures such as the Transformer \cite{vaswaniAttentionAllYou2017}, capable of incorporating contextual information, have started to be used for emotion prediction. Transformers have been  successfully used to recognize emotions with multimodal inputs composed of text, visual, audio and physiological signals \cite{tsaiMultimodalTransformerUnaligned2019, wuAttendingEmotionalNarratives2019, huangMultimodalTransformerFusion2020, caiMultimodalSentimentAnalysis2021, chienSelfassessedEmotionClassification2021}. In addition, Transformers have been used to process time-series in general \cite{liEnhancingLocalityBreaking2019, wuDeepTransformerModels2020}, and also to process uni-modal physiological signals in particular, with the aim of recognizing emotions. Arjun et al. \cite{arjunIntroducingAttentionMechanism2021} employ a variation of the Transformer, the Vision Transformer \cite{dosovitskiyImageWorth16x162020} to process EEG signals for emotion recognition, converting the EEG signals into images using continuous wavelet transform. Behinaein et al. \cite{behinaeinTransformerArchitectureStress2021} propose to detect stress from ECG signals, by using a 1D-CNN followed by a Transformer and a FCN as classifier.

Most of the approaches for measuring emotions, including those using multimodal physiological data, have relied on supervised learning, and thus are limited by the availability of labeled training data.
Using self-supervised pre-training can improve performances of a model \cite{erhanWhyDoesUnsupervised2010}, as it allows to learn more general representations, thus avoiding overfitting in the downstream task. This is especially important for tasks with limited labeled data. Sarkar and Etemad \cite{sarkarSelfSupervisedLearningECGBased2020} pre-train a 1D-CNN using a self-supervised task to learn representations from ECG signals. Their self-supervised task consists in first transforming the signal, with operations such as scaling or adding noise, and then using the network to predict which transformation has been applied. Ross et al. \cite{rossUnsupervisedMultimodalRepresentation2021} learn representations from ECG signals using auto-encoders based on 1D-CNN. In both approaches, once the representations have been learned, they are used to predict emotions.

In contrast with the two previously mentioned approaches, we propose to take into account contextual information during pre-training by using a Transformer-based model. Such an approach has been used for pre-training Transformers from visual, speech and textual modalities  \cite{macaryUseSelfSupervisedPreTrained2021, khareMultiModalEmbeddingsUsing2020, rahmanIntegratingMultimodalInformation2020, siriwardhanaMultimodalEmotionRecognition2020, khareSelfSupervisedLearningCrossModal2021}. 
Haresamudram et al. use this approach to pre-train a Transformer for human activity recognition using  accelerometer and gyroscope data
 \cite{haresamudramMaskedReconstructionBased2020}. Zerveas et al. \cite{zerveasTransformerbasedFrameworkMultivariate2021a} develop a framework for multivariate time-series representation learning, by pre-training a Transformer-based architecture. However, none of these works deal with uni-modal physiological signals. 
In this work, we have extended this approach for use with ECG signals. Specifically, we investigate the effectiveness of pre-training a Transformer for ECG emotion recognition, which to the best of our knowledge has not been done before.

\section{Our approach}

\begin{figure*}[tb]
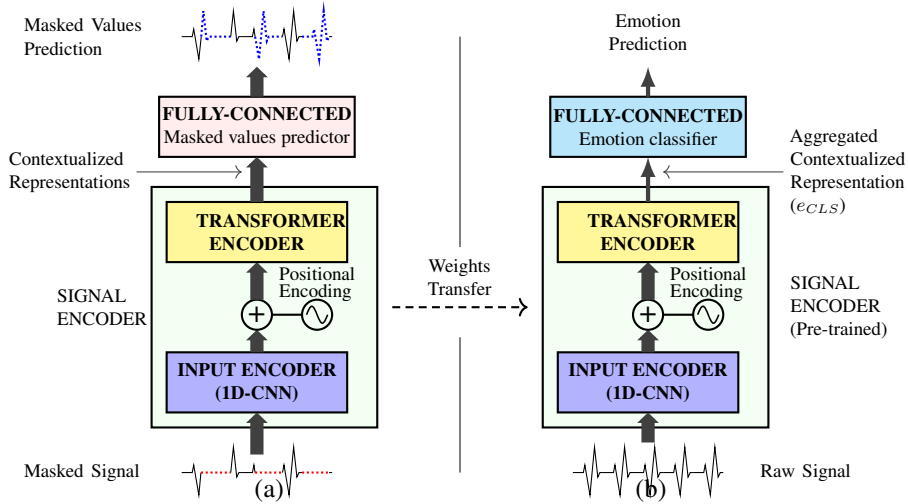

\centering
\include{tikz/approach}
\vspace*{-7.5mm}
(a) \hspace{4.5cm} (b)
\caption{Our approach with self supervised learning based on a Transformer (a) and fine-tuning strategy for learning the final emotion predictor (b).}
\label{fig:approach}
\end{figure*}

Our framework for using deep learning for emotion recognition is based on the following two steps: first, we need to obtain contextualized representations from time-series signals using a deep model; then, we use those representations to perform the targeted downstream task. In this paper, the considered physiological time-series are raw ECG signals, and the downstream task is binary emotion recognition: predicting high/low levels of arousal, and high/low levels of valence.

For the first step (see Figure \ref{fig:approach}.a), we developed a signal encoder based on deep neural networks and attention, to obtain contextualized representations from ECG signals. The main component of the signal encoder is a Transformer \cite{vaswaniAttentionAllYou2017}. This signal encoder is pre-trained with a self-supervised task, using unlabeled ECG data. For the second step (see Figure \ref{fig:approach}.b), we fine-tune the whole model (the signal encoder and the fully-connected classifier) for our downstream task of binary emotion recognition, using labeled ECG data.

In the following subsections, we describe in detail the different components of our approach.

\subsection{Learning Contextualized Representations}
At the heart of our signal encoder is a Transformer encoder \cite{vaswaniAttentionAllYou2017}, which we use to learn contextualized representations of ECG signals. In Transformers, contextual information is obtained through an attention mechanism, with the attention function considered as a mapping of a query vector along with a group of key-value vector pairs to an output. In the case of the Transformer encoder, each position in the output pays attention to all positions in the input. Several attention modules (also called \textit{heads}) are used, creating various representation subspaces and improving the ability of the model to be attentive to different positions. The Transformer encoder is constructed by stacking several layers containing a multi-head attention module followed by a fully-connected network applied to each position, with residual connections. Since our implementation of the Transformer is almost identical to the one described in \cite{vaswaniAttentionAllYou2017}, we refer the readers to this paper for further details.

\begin{figure}[tb]
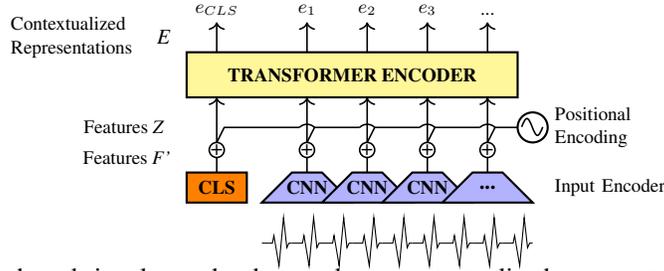

\centering
\include{tikz/signal_encoder}
\vspace*{-7.5mm}
\caption{Our Transformer-based signal encoder that produces contextualized representations. The aggregated representation $e_{CLS}$ is used for classification.}
\label{fig:signalencoder}
\end{figure}

In Figure \ref{fig:signalencoder}, we present our signal encoder, which we describe in the remainder of this subsection.

\textit{Input Encoder:} to process an ECG signal with the Transformer, we first encode it into \textit{s} feature vectors of dimension $d_{\text{model}}$ that represent each one of the $s$ values of the ECG signal. We use 1D Convolutional Neural Networks (1D-CNN) to perform this encoding, like in \cite{haresamudramMaskedReconstructionBased2020, tsaiMultimodalTransformerUnaligned2019,  baevskiWav2vecFrameworkSelfSupervised2020}. Thus, for a raw input signal $X=\{x_1,...,x_s\}$ where $x_i$ is a single value, after encoding $X$ with the input encoder we obtain features $F=\{f_1, ..., f_s\}$ where $f_i \in \mathbb{R}^{d_{\text{model}}}$.

\textit{CLS token:} given that our downstream task is a classification task, we need to obtain a single representation of the whole processed signal at the output of our signal encoder. Similar to what is done in BERT \cite{devlinBERTPretrainingDeep2019}, we append a special classification token (CLS) at the start of the feature sequence $F$, resulting in the sequence $F' = \{CLS, f_1, ..., f_s\}$. We use a trainable vector of dimension $d_\text{model}$ as CLS token. At the output of the Transformer, we obtain an embedding of the CLS token ($e_\text{CLS}$), along with the rest of the representations of the signal (see Figure \ref{fig:signalencoder} and Equation \ref{eq:transformer}). Through the attention mechanisms of the Transformer, $e_\text{CLS}$ is capable of aggregating information from the entire input signal and its contextualized representations. For this reason, at classification time, $e_{CLS}$ can be used as input for the classifier network.

\textit{Positional Encoding:} positional information of each input is required so that the Transformer can take into account the actual ordering of time-steps in the input sequence. As in \cite{vaswaniAttentionAllYou2017}, we use fixed sinusoidal positional embeddings. We sum the positional embeddings with the features $F'$:
\begin{equation} \label{eq:posencoding}
 Z = \{CLS + pe_0, f_1 + pe_1, ..., f_s+pe_s\},
\end{equation}
where $pe_i \in \mathbb{R}^{d_\text{model}}$ is the positional embedding for time-step $i$. We then apply layer normalization \cite{baLayerNormalization2016} to $Z$. Please refer to \cite{vaswaniAttentionAllYou2017} for details on how to obtain the positional embeddings.

\textit{Transformer Encoder:} we obtain contextualized representations $E$ using a Transformer encoder with $h$ heads and $l$ layers on the sequence $Z$:
\begin{equation} \label{eq:transformer}
 E = \{e_{CLS}, e_1, ...,e_s\} = \text{Transformer}_{h,l}(Z).
\end{equation}

We then use the representations $E$ for emotion recognition, as is described in Section \ref{sec:finetuning}

\subsection{Pre-training Task}
To pre-train our signal encoder, we employ a self-supervised approach inspired in BERT \cite{devlinBERTPretrainingDeep2019}. We mask random segments of a certain length by replacing them with zeros, and then we train our model to predict the masked values, as shown in Figure \ref{fig:approach}a. Labeled data is not needed for this step. 

Similar to \cite{baevskiWav2vecFrameworkSelfSupervised2020}, a proportion $p$ of points is randomly selected from the input signal as starting points for masked segments, and then for each starting point the subsequent $M$ points are masked. The masked segments may overlap.

To predict masked points, we use a fully-connected network (FCN)  on top of the signal encoder, as shown in \mbox{Figure \ref{fig:approach}a.} We only predict values of masked inputs, as opposed to reconstructing the whole signal. We use the mean square error between predicted and real values as the reconstruction loss $\mathcal{L}_{r}$ during pre-training:
\begin{equation} \label{eq:rec_loss}
 \mathcal{L}_{r} = \frac{1}{N_m}\sum_{j=1}^{N_m}(\hat{x}_j - x_{p(j)})^2,
\end{equation}
where $N_m$ is the number of masked values, $\hat{x}_j$ is the prediction corresponding to the $j^{th}$ masked value, and $x_{p(j)}$ is the original input value selected to be the $j^{th}$ masked value, whose position is $p(j)$ in the input signal.

\subsection{Fine-tuning} \label{sec:finetuning}
We fine-tune our model to perform binary emotion prediction, as shown in Figure \ref{fig:approach}b. This step is supervised, using labeled data. To make the prediction, a FCN is added on top of the signal encoder, using $e_{CLS}$ as input. We initialize the signal encoder with the weights obtained after pre-training, while the FCN is randomly initialized. We then fine-tune all the parameters of the model, including the pre-trained weights. For this task, we minimize the binary cross-entropy loss $\mathcal{L}_{ft}$:
\begin{equation} \label{eq:sup_loss}
 \mathcal{L}_{ft} = - w_p y \log[\sigma(out)] - (1-y)\log[1- \sigma(out)]
\end{equation}
where $y$ is an indicator variable with value 1 if the class of the ground truth is positive and 0 if it is negative, $out$ is the output of the classifier, $\sigma$ is the sigmoid function, and $w_p$ is the ratio of negative to positive training samples, used to compensate unbalances that may be present in the dataset.

\section{Experimental Setup}

\begin{table*}[t]
\centering
\caption{Comparison of different strategies of our approach on AMIGOS dataset}
\footnotesize
\begin{tabular}{p{3cm} p{3.5cm} M{2cm} M{2cm} M{2cm} M{2cm}} 
\toprule[1pt]
& & \textbf{Arousal Acc.} & \textbf{Arousal F1} & \textbf{Valence Acc.} & \textbf{Valence F1}\\
\midrule[1pt]
\multirow{5}{3.0cm}{\textbf{Aggregation Method}} & Max-Pooling 1 & 0.85$\pm6.6\mathrm{e}^{-3}$ & 0.84$\pm6.4\mathrm{e}^{-3}$ & 0.78$\pm6.5\mathrm{e}^{-3}$ & 0.78$\pm6.6\mathrm{e}^{-3}$  \\
& Max-Pooling 2 & 0.86$\pm7.4\mathrm{e}^{-3}$ & 0.84$\pm7.3\mathrm{e}^{-3}$ & 0.8$\pm6.3\mathrm{e}^{-3}$ & 0.8$\pm5.9\mathrm{e}^{-3}$\\
& Average-Pooling 1 & 0.87$\pm8.3\mathrm{e}^{-3}$ & \textbf{0.87$\pm7.3\mathrm{e}^{-3}$} & 0.82$\pm6.2\mathrm{e}^{-3}$ & 0.82$\pm6.7\mathrm{e}^{-3}$ \\
& Average-Pooling 2 & \textbf{0.88$\pm4.4\mathrm{e}^{-3}$} & \textbf{0.87$\pm4.6\mathrm{e}^{-3}$} & \textbf{0.83$\pm6.4\mathrm{e}^{-3}$} & \textbf{0.83$\pm6.6\mathrm{e}^{-3}$}\\
& Last Representation & 0.85$\pm1.3\mathrm{e}^{-2}$ & 0.84$\pm1.2\mathrm{e}^{-2}$ & 0.8$\pm7.6\mathrm{e}^{-3}$ & 0.8$\pm8.0\mathrm{e}^{-3}$\\
\midrule[1pt]
\multirow{2}{3.0cm}{\textbf{Segment Length}} & 40 seconds & 0.86$\pm1.2\mathrm{e}^{-2}$ & 0.85$\pm1.1\mathrm{e}^{-2}$ & 0.82$\pm1.0\mathrm{e}^{-2}$ & 0.81$\pm9.9\mathrm{e}^{-3}$\\
& 20 seconds & 0.87$\pm5.6\mathrm{e}^{-3}$ & 0.86$\pm6.4\mathrm{e}^{-3}$ & 0.82$\pm7.8\mathrm{e}^{-3}$ & 0.82$\pm8.1\mathrm{e}^{-3}$\\
\midrule[1pt]
\textbf{Our Best Approach} & CLS with 10s segment & \textbf{0.88$\pm5.4\mathrm{e}^{-3}$} & \textbf{0.87$\pm5.4\mathrm{e}^{-3}$} & \textbf{0.83$\pm7.8\mathrm{e}^{-3}$} & \textbf{0.83$\pm7.4\mathrm{e}^{-3}$} \\
\bottomrule[1pt]
\end{tabular}
\label{table:ablations}
\end{table*}

\begin{table}[t]
\centering
\caption{No Pre-training vs pre-trained model}
\footnotesize
\begin{tabular}{M{1.5cm} M{2cm} M{2cm} M{2cm} M{2cm}}
\toprule[1pt]
 \multirow{1}{1.5cm}{\textbf{Pre-train}} & \textbf{Arousal Acc.} & \textbf{Arousal F1} & \textbf{Valence Acc.} & \textbf{Valence F1}\\
\midrule[1pt]
No & 0.85$\pm5.6\mathrm{e}^{-3}$ & 0.84$\pm5.8\mathrm{e}^{-3}$ & 0.8$\pm6.5\mathrm{e}^{-3}$ & 0.8$\pm6.4\mathrm{e}^{-3}$ \\
Yes & \textbf{0.88$\pm5.4\mathrm{e}^{-3}$} & \textbf{0.87$\pm5.4\mathrm{e}^{-3}$} & \textbf{0.83$\pm7.8\mathrm{e}^{-3}$} & \textbf{0.83$\pm7.4\mathrm{e}^{-3}$} \\
\bottomrule[1pt]
\end{tabular}
\label{table:pretrain}
\end{table}

In this section, we describe the experimental choices taken to evaluate our approach for a downstream task of binary emotion recognition (high/low levels of arousal and valence), on ECG signals. We present the datasets used, the pre-processes employed, and the parametrization of our two steps of pre-training and fine-tuning.

\subsection{Datasets}
For pre-training, we only require datasets that contain ECG signals, regardless of why they were actually collected or which labeling they have, if any. The datasets that we use in our experiments are: ASCERTAIN \cite{subramanianASCERTAINEmotionPersonality2018}, DREAMER \cite{katsigiannisDREAMERDatabaseEmotion2018}, PsPM-FR \cite{tzovaraPsPMFRSCRECG2018}, PsPM-HRM5 \cite{paulusPsPMHRM5SCRECG2020}, PsPM-RRM1-2 \cite{bachPsPMRRM12SCRECG2019}, and PsPM-VIS \cite{xiaPsPMVISSCRECG2020}. We also employ the AMIGOS dataset \cite{correaAMIGOSDatasetAffect2018}, taking care of not using the same data for pre-training and evaluating our model, as this dataset is also used for the downstream task. To gather as much data as possible, we use all the ECG channels available in the datasets. For ASCERTAIN, we discard some signals according to the quality evaluation provided in the dataset: if a signal has a quality level of 3 or worse in the provided scale, it is discarded. In total, there are around 230 hours of ECG data for pre-training.

To fine-tune our model to predict emotions, we use the \mbox{AMIGOS} dataset \cite{correaAMIGOSDatasetAffect2018}. In this dataset, 40 subjects watched videos specially selected to evoke an emotion. After watching each video, a self-assessment of their emotional state is conducted. In this assessment, subjects rated their levels of arousal and valence on a scale of 1 to 9. Of the 40 subjects, 37 watched a total of 20 videos, while the other 3 subjects watched only 16 videos. During each trial, ECG data were recorded on both left and right arms. We use data only from the left arm to fine-tune our model. AMIGOS includes a pre-processed version of the data, that was down-sampled to 128Hz and filtered with a low-pass filter with 60Hz cut-off frequency. We use these pre-processed data for our experiments, including the pre-training phase. The ECG data that we use for fine-tuning amounts to around 65 hours of recordings.

\subsection{Signal Pre-processing}  We first filter signals with an 8\textsuperscript{th} order Butterworth band-pass filter, having a low-cut-off frequency of 0.8Hz and a high-cut-off frequency of 50Hz. We then down-sample the signals to 128 Hz, except for AMIGOS which already has that sampling rate. Signals are normalized so they have zero-mean and unit-variance, for each subject independently. Signals are finally divided into 10-second segments (we also report results for segments of 20 seconds and 40 seconds).

\subsection{Pre-training}
As stated previously, we use ASCERTAIN, DREAMER, PsPM-FR, PsPM-RRM1-2, PsPM-VIS, and AMIGOS for pre-training. Since we also use AMIGOS for fine-tuning, we need to avoid using the same segments both for pre-training and for evaluating the model. To do this, we pre-train two models, one using half of the data from AMIGOS, and the second using the other half. When testing our model with certain segments from AMIGOS, we fine-tune the model that was pre-trained with the half of AMIGOS that do not contain those segments. More details are given in Section \ref{sec:Setup_finetune}. In total, both of our models are pre-trained with 83401 10-second segments.

We select a proportion of $p=0.0325$ points from each input segment to be the starting point of a masked span of length $M=20$, resulting in around 47\% of the input values masked.

The input encoder is built with 3 layers of 1D-CNN with ReLU activation function. We use layer normalization \cite{baLayerNormalization2016} on the first layer, and at the output of the encoder. Kernel sizes are (65, 33, 17), the numbers of channels are (64, 128, 256) and the stride for all layers is 1. This results in a receptive field of 113 input values or 0.88s. We selected this receptive field size because it is comparable with the typical interval between peaks on an ECG signal, which is between 0.6s and 1s, including when experiencing emotions \cite{wuHowAmusementAnger2019}.

The Transformer in our signal encoder has a model dimension $d_\text{model}=256$, 2 layers and 2 attention heads, with its FCN size of $d_ {model} \cdot 4=1024$. The FCN used to predict the masked values consists of a single linear layer of size $d_\text{model}/2=128$ followed by a ReLU activation function. An additional linear layer is used to project the output vector to a single value, which corresponds to the predicted value of a masked point.

We pre-train the two models for 500 epochs, warming up the learning rate over the first 30 epochs up to a value of 0.001 and using linear decay after that. We employ Adam optimization, with $\beta_1=0.9$, $\beta_2=0.999$, and $L_2$ weight decay of 0.005. We use dropout of 0.1 at the end of the input encoder, after the positional encoding, and inside the Transformer.

We tuned the number of layers and heads in the Transformer, the learning rate, and the warm-up duration using the Ray Tune framework \cite{liawTuneResearchPlatform2018} with BOHB optimization \cite{falknerBOHBRobustEfficient2018}.

\subsection{Fine-Tuning} \label{sec:Setup_finetune}
We fine-tune our model (both the signal encoder and FCN classifier) for emotion recognition with the AMIGOS dataset, using each of the 10-second segments as a sample. As labels, we use the emotional self-assessments given in the dataset. Since these assessments provide values of arousal and valence on a scale 1 to 9, we use the average arousal and the average valence as threshold value to determine a low or a high level.

We use 10-fold cross-validation to evaluate our approach. Recall that we pre-train two signal encoders. After dividing AMIGOS into 10 folds, we use folds 1 to 5 to pre-train one signal encoder ($SE_1$), and folds 6 to 10 to pre-train the second one ($SE_2$) (and all data from the other datasets, for both). Then, when we fine-tune the models to be tested with folds 1 to 5, we use the weights from $SE_2$ to initialize the signal encoder parameters. In a similar fashion, we use $SE_1$ as initialization point of the signal encoder when we fine-tune the models to be tested with folds 6 to 10. This method allows us to pre-train, fine-tune and test our model in a more efficient way than pre-training 10 different models, one for each fold, while retaining complete separations between training and testing data.

The FCN classifier used to predict emotions has two hidden layers of sizes [1024, 512] with ReLU activation functions, and an output layer that projects the output to a single value. We fine-tune one model to predict arousal and another to predict valence. For each task, we fine-tune our model for 100 epochs using Adam optimization, with $\beta_1 = 0.9$, $\beta_2=0.999$ and $L_2$ weight decay of 0.00001. We start with a learning rate of 0.0001, and decrease it every 45 epochs by a factor of 0.65.  We keep using a dropout of 0.1 at the end of the input encoder, after the positional encoding, and inside the Transformer. We use dropout of 0.3 in the FCN classifier.

We used Ray Tune with BOHB, as we did on pre-training, to tune the learning rate, the learning rate schedule, the shape and dropout of the FCN classifier, and the $L_2$ weight decay.

\section{Results}

\begin{table*}[tb]
\centering
\caption{Comparison of different methods on AMIGOS dataset}
\noindent\makebox[\textwidth]{%
\footnotesize
\begin{tabular}{p{1.7cm} p{3.7cm} M{1.25cm} M{2cm} M{1.7cm} M{1.5cm} M{1.7cm} M{1.55cm}} 
\toprule[1pt]
& \multirow{1}{3.8cm}{\textbf{Model}} & \multirow{1}{1.5cm}{\textbf{Subj. Ind.}} & \multirow{1}{2.1cm}{\textbf{Input Seg. Size}} & \textbf{Arousal Acc}. & \textbf{Arousal F1} & \textbf{Valence Acc.} &\textbf{ Valence F1}\\
\midrule[1pt]
\multirow{5}{2.25cm}{\textbf{Various experiment protocols}} & Gaussian Naive Bayes \cite{correaAMIGOSDatasetAffect2018} & Yes & 20 seconds & - & 0.551 & - & 0.545\\
& 1D-CNN \cite{santamaria-granadosUsingDeepConvolutional2019} & No & 200 peaks & 0.81 & 0.76 & 0.71 & 0.68\\
& 2D-CNN \cite{siddharthUtilizingDeepLearning2019} & Yes & Not segmented & 0.83 & 0.76 & 0.82 & 0.80\\
& 1D-CNN with LSTM \cite{harperBayesianDeepLearning2020} & Yes & Not segmented & - & - & 0.81 & 0.80\\
& \multirow{1}{4.1cm}{Convolutional autoencoder \cite{rossUnsupervisedMultimodalRepresentation2021}} & No & 10 seconds & 0.85 & 0.89 & - & -\\
\midrule[1pt]
\multirow{2}{2cm}{\textbf{Our protocol}} & Pre-trained CNN \cite{sarkarSelfSupervisedLearningECGBased2020} & No & 10 seconds & 0.85$\pm5.4\mathrm{e}^{-3}$ & 0.84$\pm5.3\mathrm{e}^{-3}$ & 0.77$\pm5.5\mathrm{e}^{-3}$ & 0.77$\pm5.1\mathrm{e}^{-3}$\\
& \multirow{1}{4.3cm}{\textbf{Pre-trained Transformer (ours)}} & No & 10 seconds & \textbf{0.88$\pm5.4\mathrm{e}^{-3}$} & \textbf{0.87$\pm5.4\mathrm{e}^{-3}$} & \textbf{0.83$\pm7.8\mathrm{e}^{-3}$} & \textbf{0.83$\pm7.4\mathrm{e}^{-3}$}\\
\bottomrule[1pt]
\end{tabular}
}
\label{table:allsota}
\end{table*}

In our results, we use as metrics the mean accuracy and mean F1-score between positive and negative classes. We report the mean and confidence intervals of the metrics across our 10 folds of cross-validation. The confidence intervals are calculated using a t-distribution with 9 degrees of freedom, for a two-sided 95\% confidence.

\subsection{Comparing Aggregation Methods and Segment Lengths}

We report in Table \ref{table:ablations} the performances of our approach for different strategical choices. Firstly, we compare different aggregation approaches to combine the contextualized representations at the output of the signal encoder, given to the FCN classifier. Secondly, we compare performances for different segment lengths used to divide the input signals.

\textit{Aggregation Method:} we compared 4 strategies for aggregating representations, to be given as input to the FCN: max-pooling, average-pooling, using only the last representation $e_s$, and using only the embedding of the CLS token $e_{CLS}$ (we call this strategy CLS). Max-pooling 1 and \mbox{average-pooling 1} are the result of max-pooling and average-pooling across all representations, to obtain a single representation of size $d_{model}=256$. Max-pooling 2 was optimized on the validation set: representations are reduced to a size of 64, divided into two groups, then max-pooling was applied on each group and the results concatenated to obtain a single representation of size 128. Average-pooling 2 was optimized on the validation set: representations are divided into 4 groups, average-pooling is applied on each group and the results concatenated to obtain a single representation of size 1024. 

We see in Table \ref{table:ablations} that the best results were obtained with average-pooling strategies and with CLS, with accuracies up to $0.88$ for arousal, for example. In the following experiments, we will thus use CLS as our aggregation method. Indeed, although results are practically identical for CLS and \mbox{average-pooling 2} (e.g. 0.88$\pm5.4\mathrm{e}^{-3}$ compared to 0.88$\pm4.4\mathrm{e}^{-3}$ accuracies for arousal), CLS has the advantage of being a commonly-used strategy for Transformers, which does not require any kind of tuning on validation data, contrary to average-pooling 2.

\textit{Segment length:} we compare 3 different segment lengths for dividing ECG signals into input instances: 10, 20, and 40 second segments. We can see in Table \ref{table:ablations}  that shorter segments lead to better results on average, both for arousal and valence. For example for arousal, 10-second segments lead to an accuracy of 0.88$\pm5.4\mathrm{e}^{-3}$, compared to 0.87$\pm5.6\mathrm{e}^{-3}$ for 20-second segments, and 0.86$\pm1.2\mathrm{e}^{-2}$ for 40-second segments. 

Two explanations emerge for this observation: firstly, since emotions are relatively volatile states, longer segmentation might cover fluctuating emotional states, thus making it harder to characterize emotion; secondly, longer segments should require more complex models (i.e. bigger Transformer and FCN), which are harder to train with the relatively restricted amount of labeled data in AMIGOS. Moreover, shorter segments are faster to process, allowing a high number of training epochs and smaller learning rates. In the following experiments, we will thus use 10-second segments.

\subsection{Effectiveness of Pre-training}

To demonstrate the effectiveness of our pre-training approach, we tested our architecture by fine-tuning our model on AMIGOS with all parameters randomly initialized, instead of using a pre-trained signal encoder (thus skipping step (a) of our process in Figure \ref{fig:approach}). As reported in Table \ref{table:pretrain}, the pre-trained model is on average significantly better than the model with no pre-training, for both accuracy and F1-score. For example, for arousal, the pre-trained model reaches an average accuracy of $0.88\pm5.4\mathrm{e}^{-3}$, compared to $0.85\pm5.6\mathrm{e}^{-3}$ for the model with no pre-training. These results illustrate the benefits of pre-training Transformers for our task. Moreover, during our experiments, we observed that the model with no pre-training had a tendency to overfit quickly, which was not the case for the pre-trained model. Pre-training the model on many different datasets should increase its robustness to overfitting when fine-tuning on a specific dataset.

\subsection{Comparisons With Other Approaches}

We report in Table \ref{table:allsota} various state-of-the-art results for emotion recognition from ECG signals on the AMIGOS dataset. The first section of the table contains results from works which all use different experiment protocols, such as different segment sizes, different separations of data into training and test sets, subject dependent and independent evaluations, etc. These results are therefore not directly comparable with one another, nor are they directly comparable with ours. Nevertheless, we report them to showcase the variety of state-of-the-art approaches published for this task, and give a relative idea of achieved performances on AMIGOS.

To compare our approach with another state-of-the-art approach as fairly as possible, it is required that both use exactly the same experiment protocol. For this, we fully retrained and tested the pre-trained CNN approach proposed by Sarkar and Etemad \cite{sarkarSelfSupervisedLearningECGBased2020}, with the experiment protocol we presented. To this end, we use the implementation provided by the authors\footnote{https://code.engineering.queensu.ca/pritam/SSL-ECG}. To ensure fair comparisons, the exact same data was used to pre-train, fine-tune, and test both our approach and also Sarkar and Etemad’s approach, for each fold of cross-validation.

We see in Table \ref{table:allsota} that our approach achieves better performance on average than Sarkar and Etemad’s approach with the same experiment protocol, for both arousal and valence. For example, our approach achieves an F1-score of 0.83$\pm7.4\mathrm{e}^{-3}$ for valence, compared to 0.77$\pm5.1\mathrm{e}^{-3}$ for the pre-trained CNN. These results are statistically significant with $p < 0.01$ following a t-test.

This final set of results shows that our approach, and more generally self-supervised Transformer-based approaches, can be successfully applied to obtain contextualized representations from ECG signals for emotion recognition tasks.

\section{Conclusions and Perspectives}

In this paper, we investigate the use of transformers for recognizing arousal and valence from ECG signals. 
This approach used self-supervised learning for pre-training from unlabeled data, followed by fine-tuning with labeled data. Our experiments indicate that the model builds robust features for predicting arousal and valence on the AMIGOS dataset, and provides very promising results in comparison to  recent state-of-the-art methods. This work showcases that self-supervision and attention-based models such as Transformers can be successfully used for research in affective computing.

Multiple perspectives emerge from our work. New pre-training tasks can be investigated: other methods such as contrastive loss or triplet loss might be more efficient with regards to the specificities of ECG signals, compared to masked points prediction which we used in this work. Extending our work to other input modalities (EEC, GSR, and even non-physiological inputs such as ambient sensors) and, in general, to process multimodal situations could prove useful for improving performances of emotion recognition. Finally, larger scale experiments, with new datasets captured in varied situations, will allow for a better understanding of the behaviour of our approach.

\vspace{0.25cm}
\textbf{Acknowledgements:} This work has been partially supported by the MIAI Multidisciplinary AI Institute at the Univ. Grenoble Alpes:  (MIAI@Grenoble Alpes - ANR-19-P3IA-0003).

\bibliographystyle{unsrt}
\bibliography{refs}

\end{document}

%% file: tikz/approach.tex
\begin{tikzpicture}[inner sep=3pt, thick,scale=0.4, every node/.style={scale=0.8}]

\begin{scope}[shift={(0,2.5)}]
\draw [thin] (0, 0) -- (0.35, 0) -- (0.40, 0.25) -- (0.57, -0.75) -- (0.64, 0);
\draw [red, densely dotted] (0.64, 0) -- (1.63, 0);
\draw [thin] (1.63, 0) -- (1.72, 0.85) -- (1.83, -0.15) -- (1.9, 0) -- (2, 0);
\draw [thin] (2, 0) -- (2.35, 0) -- (2.40, 0.25) -- (2.44, 0);
\draw [red, densely dotted] (2.44, 0) -- (3.35, 0);
\draw [thin] (3.35, 0) -- (3.40, 0.25) -- (3.57, -0.75) -- (3.72, 0.85) -- (3.83, -0.15) -- (3.9, 0) -- (4, 0);
\draw [red, densely dotted] (4, 0) -- (4.9, 0);
\draw [thin] (4.9, 0) -- (5, 0);
\end{scope}

\begin{scope}[shift={(0, 4)}]
\draw [fill=green!05] (-1, 0) rectangle (6.5, 8);
\node[outer sep=10pt] (anc_transfo) at (11,1) {};
\node[outer sep=3pt, inner sep=0pt] (tr_l) at (0,0) {};
\node[outer sep=3pt, inner sep=0pt] (tr_h) at (0,2) {};
\end{scope}

\begin{scope}[shift={(-0.5, 4.5)}]
\draw [fill=blue!30] (0, 0) rectangle (6, 2) node[pos=0.5, align=center] {\footnotesize \textbf{INPUT ENCODER} \\ \footnotesize \textbf{(1D-CNN)}};
\node[outer sep=10pt] (anc_transfo) at (11,1) {};
\node[outer sep=3pt, inner sep=0pt] (tr_l) at (0,0) {};
\node[outer sep=3pt, inner sep=0pt] (tr_h) at (0,2) {};
\end{scope}

\begin{scope}[shift={(2.5,6.5)}]
\draw[-{Triangle[width=8pt,length=4pt]}, line width=5pt, darkgray] (0,0) -- (0, 0.75);
\draw (0, 1.25) circle [radius=0.5] node (plus) {\Large +};
\draw (0.5, 1.25) -- (1.5, 1.25);
\begin{scope}[shift={(2,1.25)}]
    \draw (0,0) circle [radius=0.5];

    \begin{scope}[shift={(-0.35,0)}]
        \draw [semithick, scale = 0.7, yscale = 0.8] (0, 0) sin (0.25, 0.5) cos (0.5, 0) sin (0.75, -0.5) cos (1, 0);
    \end{scope}

    \begin{scope}[shift={(-1.5, 1.35)}]
    \node [right, align=left] at (0,0) {\footnotesize Positional};
    \node [right, align=left] at (0,-.6) {\footnotesize Encoding};
    \end{scope}    

\end{scope}
\draw[-{Triangle[width=8pt,length=4pt]}, line width=5pt, darkgray] (plus) -- (0, 3.0);
\end{scope}

\begin{scope}[shift={(-0.5, 9.5)}]
\draw [fill=yellow!50, text width=2cm] (0, 0) rectangle (6, 2) node[pos=0.5, align=center] {\footnotesize \textbf{TRANSFORMER ENCODER}};
\node[outer sep=10pt] (anc_transfo) at (11,1) {};
\node[outer sep=3pt, inner sep=0pt] (tr_l) at (0,0) {};
\node[outer sep=3pt, inner sep=0pt] (tr_h) at (0,2) {};
\end{scope}

\begin{scope}[shift={(0, 13)}]
\draw [fill=pink!30] (-.75, 0) rectangle (5.75, 2) node[pos=0.5, align=center] {\footnotesize \textbf{FULLY-CONNECTED} \\ \footnotesize Masked values predictor};
\node[outer sep=10pt] (anc_transfo) at (11,1) {};
\node[outer sep=3pt, inner sep=0pt] (tr_l) at (0,0) {};
\node[outer sep=3pt, inner sep=0pt] (tr_h) at (0,2) {};
\end{scope}

\begin{scope}[shift={(0,17)}]
\draw [thin] (0, 0) -- (0.35, 0) -- (0.40, 0.25) -- (0.57, -0.75) -- (0.65, 0);
\draw [blue, densely dotted] (0.65, 0) -- (0.72, 0.85) -- (0.83, -0.15) -- (0.9, 0) -- (1.63, 0);
\draw [thin] (1.63, 0) -- (1.72, 0.85) -- (1.83, -0.15) -- (1.9, 0) -- (2, 0);
\draw [thin] (2, 0) -- (2.35, 0) -- (2.40, 0.25) -- (2.44, 0);
\draw [blue, densely dotted] (2.44, 0) -- (2.57, -0.75) -- (2.72, 0.85) -- (2.83, -0.15) -- (2.9, 0) -- (3.35, 0);
\draw [thin] (3.35, 0) -- (3.40, 0.25) -- (3.57, -0.75) -- (3.72, 0.85) -- (3.83, -0.15) -- (3.9, 0) -- (4, 0);
\draw [blue, densely dotted] (4, 0) -- (4.35, 0) -- (4.40, 0.25) -- (4.57, -0.75) -- (4.72, 0.85) -- (4.83, -0.15) -- (4.9, 0);
\draw [thin] (4.9, 0) -- (5, 0);
\end{scope}

\draw[-{Triangle[width=8pt,length=4pt]}, line width=5pt, darkgray] (2.5,3.1) -- (2.5, 4.5);
\draw[-{Triangle[width=8pt,length=4pt]}, line width=5pt, darkgray] (2.5,11.5) -- (2.5, 13);
\draw[-{Triangle[width=8pt,length=4pt]}, line width=5pt, darkgray] (2.5,15) -- (2.5, 16);

\node [left, text width=2.5cm] at (0, 2.5) {\small Masked Signal};
\node [left, text width=1.7cm] at (-0.5, 8) {\small SIGNAL ENCODER};
\node [left, text width=2.5cm] at (-0.5, 12.5) {\small Contextualized Representations};
\draw [->, darkgray, thin] (-1.5, 12.5) -- (2, 12.5);
\node [left, text width=2.5cm] at (0, 17) {\small Masked Values Prediction};

\node [text width=2cm, align=center] at (9.25, 9) {\small Weights Transfer};
\draw[->, densely dashed] (7,8) -- (11.5,8);
\draw[very thin] (9.25,2.5) -- (9.25,7);
\draw[very thin] (9.25,10) -- (9.25,17);

\begin{scope}[shift={(13,0)}]

\begin{scope}[shift={(0,2.5)}]
\draw (0, 0)[thin] -- ++(0.35, 0) -- ++(0.05, 0.25) -- ++ (0.17, -1) -- ++(0.15, 1.6) -- ++(0.11, -1) -- ++(0.07, 0.15) -- ++(0.1, 0);
\draw (1, 0)[thin] -- ++(0.35, 0) -- ++(0.05, 0.25) -- ++ (0.17, -1) -- ++(0.15, 1.6) -- ++(0.11, -1) -- ++(0.07, 0.15) -- ++(0.1, 0);
\draw (2, 0)[thin] -- ++(0.35, 0) -- ++(0.05, 0.25) -- ++ (0.17, -1) -- ++(0.15, 1.6) -- ++(0.11, -1) -- ++(0.07, 0.15) -- ++(0.1, 0);
\draw (3, 0)[thin] -- ++(0.35, 0) -- ++(0.05, 0.25) -- ++ (0.17, -1) -- ++(0.15, 1.6) -- ++(0.11, -1) -- ++(0.07, 0.15) -- ++(0.1, 0);
\draw (4, 0)[thin] -- ++(0.35, 0) -- ++(0.05, 0.25) -- ++ (0.17, -1) -- ++(0.15, 1.6) -- ++(0.11, -1) -- ++(0.07, 0.15) -- ++(0.1, 0);
\end{scope}

\begin{scope}[shift={(0, 4)}]
\draw [fill=green!05] (-1, 0) rectangle (6.5, 8);
\node[outer sep=10pt] (anc_transfo) at (11,1) {};
\node[outer sep=3pt, inner sep=0pt] (tr_l) at (0,0) {};
\node[outer sep=3pt, inner sep=0pt] (tr_h) at (0,2) {};
\end{scope}

\begin{scope}[shift={(-0.5, 4.5)}]
\draw [fill=blue!30] (0, 0) rectangle (6, 2) node[pos=0.5, align=center] {\footnotesize \textbf{INPUT ENCODER} \\ \footnotesize \textbf{(1D-CNN)}};
\node[outer sep=10pt] (anc_transfo) at (11,1) {};
\node[outer sep=3pt, inner sep=0pt] (tr_l) at (0,0) {};
\node[outer sep=3pt, inner sep=0pt] (tr_h) at (0,2) {};
\end{scope}

\begin{scope}[shift={(2.5,6.5)}]
\draw[-{Triangle[width=8pt,length=4pt]}, line width=5pt, darkgray] (0,0) -- (0, 0.75);
\draw (0, 1.25) circle [radius=0.5] node (plus) {\Large +};
\draw (0.5, 1.25) -- (1.5, 1.25);
\begin{scope}[shift={(2,1.25)}]
    \draw (0,0) circle [radius=0.5];

    \begin{scope}[shift={(-0.35,0)}]
        \draw [semithick, scale = 0.7, yscale = 0.8] (0, 0) sin (0.25, 0.5) cos (0.5, 0) sin (0.75, -0.5) cos (1, 0);
    \end{scope}

    \begin{scope}[shift={(-1.5, 1.35)}]
    \node [right, align=left] at (0,0) {\footnotesize Positional};
    \node [right, align=left] at (0,-.6) {\footnotesize Encoding};
    \end{scope}    

\end{scope}
\draw[-{Triangle[width=8pt,length=4pt]}, line width=5pt, darkgray] (plus) -- (0, 3.0);
\end{scope}

\begin{scope}[shift={(-0.5, 9.5)}]
\draw [fill=yellow!50, text width=2cm] (0, 0) rectangle (6, 2) node[pos=0.5, align=center] {\footnotesize \textbf{TRANSFORMER ENCODER}};
\node[outer sep=10pt] (anc_transfo) at (11,1) {};
\node[outer sep=3pt, inner sep=0pt] (tr_l) at (0,0) {};
\node[outer sep=3pt, inner sep=0pt] (tr_h) at (0,2) {};
\end{scope}

\begin{scope}[shift={(0, 13)}]
\draw [fill=cyan!25] (-.75, 0) rectangle (5.75, 2) node[pos=0.5, align=center] {\footnotesize \textbf{FULLY-CONNECTED} \\ \footnotesize Emotion classifier};
\node[outer sep=10pt] (anc_transfo) at (11,1) {};
\node[outer sep=3pt, inner sep=0pt] (tr_l) at (0,0) {};
\node[outer sep=3pt, inner sep=0pt] (tr_h) at (0,2) {};
\end{scope}

\draw[-{Triangle[width=8pt,length=4pt]}, line width=5pt, darkgray] (2.5,3.5) -- (2.5, 4.5);
\draw[-{latex},  ultra thick, darkgray] (2.5,11.5) -- (2.5, 13);
\draw[-{latex}, ultra thick, darkgray] (2.5,15) -- (2.5, 16);

\node [right, text width=2.5cm] at (6, 2.5) {\small Raw Signal};
\node [right, text width=1.8cm] at (7, 8) {\small SIGNAL ENCODER (Pre-trained)};
\node [right, text width=2cm] at (7, 12.5) {\small Aggregated Contextualized Representation ($e_{CLS}$)};
\draw [->, darkgray, thin] (7, 12.5) -- (3, 12.5);
\node [below, text width=1.4cm, align=center] at (2.5, 18) {\small Emotion Prediction};

\end{scope}

\end{tikzpicture}

%% file: tikz/signal_encoder.tex
\begin{tikzpicture}[inner sep=3pt, thick,scale=0.4, every node/.style={scale=0.8}]

\begin{scope}[shift={(2.5,1.65)}]
\draw (0, 0)[thin] -- ++(0.35, 0) -- ++(0.05, 0.25) -- ++ (0.17, -1) -- ++(0.15, 1.6) -- ++(0.11, -1) -- ++(0.07, 0.15) -- ++(0.1, 0);
\draw (1, 0)[thin] -- ++(0.35, 0) -- ++(0.05, 0.25) -- ++ (0.17, -1) -- ++(0.15, 1.6) -- ++(0.11, -1) -- ++(0.07, 0.15) -- ++(0.1, 0);
\draw (2, 0)[thin] -- ++(0.35, 0) -- ++(0.05, 0.25) -- ++ (0.17, -1) -- ++(0.15, 1.6) -- ++(0.11, -1) -- ++(0.07, 0.15) -- ++(0.1, 0);
\draw (3, 0)[thin] -- ++(0.35, 0) -- ++(0.05, 0.25) -- ++ (0.17, -1) -- ++(0.15, 1.6) -- ++(0.11, -1) -- ++(0.07, 0.15) -- ++(0.1, 0);
\draw (4, 0)[thin] -- ++(0.35, 0) -- ++(0.05, 0.25) -- ++ (0.17, -1) -- ++(0.15, 1.6) -- ++(0.11, -1) -- ++(0.07, 0.15) -- ++(0.1, 0);
\draw (5, 0)[thin] -- ++(0.35, 0) -- ++(0.05, 0.25) -- ++ (0.17, -1) -- ++(0.15, 1.6) -- ++(0.11, -1) -- ++(0.07, 0.15) -- ++(0.1, 0);
\draw (6, 0)[thin] -- ++(0.35, 0) -- ++(0.05, 0.25) -- ++ (0.17, -1) -- ++(0.15, 1.6) -- ++(0.11, -1) -- ++(0.07, 0.15) -- ++(0.1, 0);
\draw (7, 0)[thin] -- ++(0.35, 0) -- ++(0.05, 0.25) -- ++ (0.17, -1) -- ++(0.15, 1.6) -- ++(0.11, -1) -- ++(0.07, 0.15) -- ++(0.1, 0);
\draw (8, 0)[thin] -- ++(0.35, 0) -- ++(0.05, 0.25) -- ++ (0.17, -1) -- ++(0.15, 1.6) -- ++(0.11, -1) -- ++(0.07, 0.15) -- ++(0.1, 0);
\end{scope}

\begin{scope}[shift={(0,3)}]
\draw [fill=orange] (0,0) rectangle (2,1) node at(1,0.5) {\footnotesize \textbf{CLS}} ;
\begin{scope}[shift={(0.5,0)}]
\def\cnn{-- +(3, 0) -- +(2, 1) -- +(1, 1) -- cycle}
\foreach \x in {2, 4,...,6}{
    \draw [black, fill=blue!30] (\x, 0) \cnn;
    \node at (\x + 1.5, 0.5) {\footnotesize \textbf{CNN}};
}
\draw [black, fill=blue!30] (8, 0) \cnn;
\node at (8 + 1.5, 0.5) {\textbf{...}};
\node [right, text width=2cm] at (11.5,0.5) {\small Input Encoder};
\end{scope}
\end{scope}

\begin{scope}[shift={(0.0,4)}]
\node [left] at (-0.3,0.5) {\small Features \textit{F'}};
\draw [semithick] (1, 0) -- +(0, 0.5);
\begin{scope}[shift={(2.0,0)}]
\foreach \x in {2, 4,...,8}{
    \draw [semithick] (\x, 0) -- +(0, 0.5);
}
\end{scope}
\end{scope}

\begin{scope}[shift={(0.0,4.5)}, out=45,in=135]
\draw [semithick] (1,0.25) circle [radius=0.25] node {+};
\draw [semithick] (1, 0.5) -- +(0.25, 0.5) -- +(2.75, 0.5);
\draw [semithick] (3.75,1) to (4.25,1);
\begin{scope}[shift={(2.0,0)}]
\foreach \x in {2, 4,...,6}{
    \draw [semithick] (\x,0.25) circle [radius=0.25] node {+};
    \draw [semithick] (\x, 0.5) -- +(0.25, 0.5) -- +(1.75, 0.5);
    \draw [semithick] (\x +1.75,1) to +(0.5,0);
}
\draw [semithick] (8,0.25) circle [radius=0.25] node {+};
\draw [semithick] (8, 0.5) -- +(0.25, 0.5) -- +(1, 0.5);
\draw (9.5,1) circle [radius=0.5];
\begin{scope}[scale = 0.7, shift={(4.05,0.65)}, yscale = 0.8]
\draw [semithick] (9,1) sin (9.25,1.5) cos (9.5,1) sin (9.75,0.5) cos (10,1);
\end{scope}
\node [right, text width=1.75cm] at (10,1) {\small Positional Encoding};
\end{scope}
\end{scope}

\begin{scope}[shift={(0,5)}]
\node [left] at (-0.5,0.5) {\small Features \textit{Z}};
\draw [->, semithick](1, 0) -- +(0, 1.5);
\begin{scope}[shift={(2.0,0)}]
\foreach \x in {2, 4,...,8}{
    \draw [->, semithick](\x, 0) -- +(0, 1.5);
}
\end{scope}
\end{scope}

\begin{scope}[shift={(0, 6.5)}]
\draw (0, 0)[black,fill=yellow!50] rectangle (11, 1.5) node[pos=.5] {\footnotesize \textbf{TRANSFORMER ENCODER}};
\node[outer sep=10pt] (anc_transfo) at (11,1) {};
\node[outer sep=3pt, inner sep=0pt] (tr_l) at (0,0) {};
\node[outer sep=3pt, inner sep=0pt] (tr_h) at (0,2) {};
\end{scope}

\begin{scope}[shift={(0, 8)}]
\node [left, text width=2.4cm] at (-0.8,0.5) {\small Contextualized Representations};
\node [left] at (-0.3,0.5) {\textit{E}};
\draw [->, semithick](1, 0) -- +(0, 1) node[above] {\small $e_{CLS}$};
\begin{scope}[shift={(2.0,0)}]
\draw [->, semithick](2, 0) -- +(0, 1) node[above] {\small $e_1$};
\draw [->, semithick](4, 0) -- +(0, 1) node[above] {\small $e_2$};
\draw [->, semithick](6, 0) -- +(0, 1) node[above] {\small $e_3$};
\draw [->, semithick](8, 0) -- +(0, 1) node[above] {\small ...};
\end{scope}
\end{scope}

\end{tikzpicture}